\documentclass[10pt,conference]{IEEEtran}

\usepackage{amsmath}
\usepackage{booktabs}
\usepackage{graphicx}
\usepackage{hyperref}
\usepackage{xcolor}
\hypersetup{hidelinks}

\title{Scaling P{\^{e}}lagos: Multi-GPU Orchestration
for Relay Belief-Propagation Decoding}

\author{\IEEEauthorblockN{Xulong Tang\IEEEauthorrefmark{1}, Dionysios Diamantopoulos\IEEEauthorrefmark{2}, Tristan M{\"u}ller\IEEEauthorrefmark{2}, Apo Kayi\IEEEauthorrefmark{2}, Nelson Mimura Gonzalez\IEEEauthorrefmark{2},\\
Muhammad Osama\IEEEauthorrefmark{1}, Seetharami Seelam\IEEEauthorrefmark{2}, Haralampos Pozidis\IEEEauthorrefmark{2},
Blake R. Johnson\IEEEauthorrefmark{2}, and Yasuko Eckert\IEEEauthorrefmark{1}}
\IEEEauthorblockA{\IEEEauthorrefmark{1}Advanced Micro Devices, Inc., Research and Advanced Development; \quad \IEEEauthorrefmark{2}IBM Research}}

\begin{document}
\bstctlcite{paper-bib-control}
\maketitle

\begin{abstract}

Fault-tolerant quantum computing requires quantum error-correction (QEC) decoders that achieve low logical error rates under real-time operating constraints. These constraints motivate heuristic decoding algorithms whose error-correction performance must be evaluated through large-scale simulation. High-throughput decoding accelerates this evaluation and the exploration of decoder parameters. Using multiple graphics processing units (GPUs) is a natural way to scale these simulations, but naively distributing syndrome batches introduces host-side preprocessing and output-collection overheads that limit scaling. To address these limitations, we extend the P{\^{e}}lagos relay belief-propagation (Relay-BP) decoder with process-isolated orchestration, parallel host preprocessing, and shared-memory output. We evaluate both one-gross ($[[144,12,12]]$) and two-gross ($[[288,12,18]]$) codes in a multi-GPU environment. Specifically, for the one-gross circuit at $R=50$, eight
\mbox{AMD Instinct\texttrademark{} MI300X GPUs} and eight AMD Instinct MI355X
GPUs achieve optimized end-to-end speedups of $5.68\times$ and
$6.23\times$, respectively. For the two-gross circuit, the corresponding speedups are
$6.92\times$ and $7.01\times$.
When throughput is calculated using GPU decoding time, the corresponding
two-gross speedups are $7.58\times$ and $7.45\times$, indicating near-linear
GPU scaling.
We deploy P{\^{e}}lagos on Lux, Oak Ridge National Laboratory's AMD Instinct MI355X-based supercomputer, as a pathfinding
application for high-throughput QEC simulation. This deployment provides a
concrete quantum-computing use case for the AI + HPC infrastructure
supporting the Genesis Mission's scientific-discovery goals.
These end-to-end measurements
demonstrate a practical path to accelerating large-scale QEC simulation
through efficient multi-GPU decoding.
\end{abstract}


\section{Introduction}
\label{sec:introduction}
Fault-tolerant quantum computers will need logical qubits with error rates
around $10^{-12}$ to run algorithms with billions of logical
operations~\cite{beverland2025failfast}. Quantum low-density parity-check
codes, such as bivariate bicycle codes~\cite{bravyi2024bb}, promise to reach
this regime with far fewer physical qubits than the surface code, but only if
their decoders are both accurate and fast. Real-time constraints favour
heuristic message-passing decoders such as relay belief-propagation
(Relay-BP)~\cite{muller2025relay},
whose performance under circuit-level noise cannot be derived analytically
and must be established by simulation.

This simulation is expensive for two compounding reasons. First, each
operating point needs many decoded syndromes. Direct Monte Carlo sampling
needs on the order of $1/\varepsilon$ decoded syndromes to resolve a logical
error rate $\varepsilon$, which limits it in practice to rates around
$10^{-6}$ and above~\cite{mayer2025rare}. Lower rates require rare-event
techniques, such as failure-spectrum fitting and splitting
methods~\cite{bravyi2013rare,mayer2025rare,beverland2025failfast}, which
replace one infeasible sampling run with many decoder evaluations on sampled
fault configurations of each weight. Second, decoder development needs many
operating points. Codes, distances, noise strengths, and decoder parameters
such as relay depth and memory-strength intervals must be swept, both for
logical memory and for the logical operations of a full instruction
set~\cite{yoder2025tourdegross}. Every point requires a separate set of simulation runs, so
decoding throughput bounds how much of this design space can be explored.
P{\^{e}}lagos~\cite{pelagos-qsw2026,muller2025relay} maps Relay-BP to portable, tuned GPU
kernels for high-throughput QEC decoding. The P{\^{e}}lagos paper introduces the
decoder's sparse message-passing kernels, workload-adaptive layouts,
device-resident loop, and cross-platform configuration cache. It also
partitions batches across devices and uses a distributed runtime for
multi-node simulation studies.
In a 312-experiment
study on 1,024 GPUs, this runtime reduced an estimated ${\sim}45$ years of
single-GPU decoding to under 20 days~\cite{pelagos-qsw2026}.
In this paper, we target a complementary aspect: how to scale syndrome processing when one host coordinates multiple (e.g., eight) GPUs.

We refer to the contiguous range of the syndrome batch that one GPU decodes as
a shard.
Although syndrome shards are independent in P{\^{e}}lagos~\cite{pelagos-qsw2026,muller2025relay}, the existing path creates
multiple GPU decoders in one Python process and dispatches work through host
threads. This organization involves three potential overheads.
First, one process owns multiple Heterogeneous-compute Interface for
Portability (HIP) contexts and coordinates their runtimes. Second, host-side syndrome and observable preprocessing overhead grows with the total batch size across all GPUs. Third, gathering correction outputs across GPUs also increases serialization and copying overhead.

In this paper, we address these deficiencies jointly with one process per GPU,
parallel host preprocessing, and shared-memory output. At relay depth
$R=50$, the integrated design achieves eight-GPU end-to-end speedups
of $5.68\times$ on \mbox{AMD Instinct\texttrademark{}} MI300X and $6.23\times$ on AMD Instinct MI355X for the one-gross
($[[144,12,12]]$) circuit, and $6.92\times$ and $7.01\times$ for the two-gross
($[[288,12,18]]$) circuit. 
For the two-gross circuit, throughput calculated using GPU decoding time
scales by $7.58\times$ on MI300X and $7.45\times$ on MI355X. Host preparation,
dispatch, and result handling limit end-to-end scaling.
We summarize the contributions as follows:
\begin{itemize}
  \item We parallelize P{\^{e}}lagos across multi-GPU platforms and achieve scaling speedup through i) one isolated process per GPU, ii) concurrent host-side shard preparation, and iii) shared memory for returning correction arrays.
  \item We evaluate on AMD multi-GPU platforms (both MI300X and MI355X GPUs) for both the one-gross and two-gross circuits.
  Compared with the original P{\^{e}}lagos, our design improves scaling as GPU count increases.
  \item We deploy and evaluate P{\^{e}}lagos as a pathfinding QEC application
  on Lux, ORNL's MI355X-based supercomputer. By demonstrating
  scalable syndrome processing, we provide a concrete use case for extending
  AI + HPC infrastructure to quantum-computing research, aligned with the
  Genesis Mission's goal of accelerating scientific
  discovery~\cite{olcf-lux,amd-lux-science}.
\end{itemize}

The rest of the paper is organized as follows. 
Section~\ref{sec:prior} reviews Relay-BP and the prior P{\^{e}}lagos implementation, and
Section~\ref{sec:orchestration} presents the orchestration design.
Section~\ref{sec:results} presents the evaluation methodology and results,
Section~\ref{sec:related-work} reviews related work, and
Section~\ref{sec:conclusion} concludes the paper.

\section{Background}
\label{sec:prior}
\subsection{Relay-BP Introduction}
\label{sec:relay-background}
Relay-BP is a message-passing decoder for quantum low-density parity check (LDPC)
codes~\cite{muller2025relay}, and is an adaptation of  the belief propagation (BP)
developed for decoding classical LDPC codes. It operates on
a decoding graph, or detector error model, that connects error mechanisms to the
detectors they trigger. Min-sum BP exchanges log-likelihood messages along
these edges and tests whether the resulting error estimate reproduces the
observed syndrome. Syndrome agreement establishes a valid candidate, but does
not guarantee a correct logical outcome. On quantum codes, loops and
degeneracy can produce oscillations and symmetric trapping sets that prevent
BP from converging.

Relay-BP introduces disordered memory to address these failures. Each error
node mixes its fixed prior with its previous marginal using a node-specific
memory strength $\gamma_j$. Varying these strengths breaks symmetry; intervals
including negative values help the decoder escape trapping sets
\cite{muller2025relay}.

Successive BP runs form relay legs, with each leg initialized from the
previous leg's final marginals. The first uses a shared $\gamma_0$; later legs
sample fresh strengths from a problem-specific interval. Decoding stops at
the leg limit $R$ or after finding $S$ valid candidates, and returns the
candidate with the lowest prior-weighted error cost~\cite{muller2025relay}.
The resulting algorithm achieves low error rates without requiring a
second-stage decoder such as ordered statistics decoding (OSD)~\cite{Panteleev_2021, Roffe2020}.
For our $S=1$ workload, each syndrome stops at its first valid candidate,
making shard completion times data-dependent.

\subsection{P{\^{e}}lagos Introduction}
\label{sec:pelagos-background}
P{\^{e}}lagos implements Relay-BP on GPUs for high-throughput QEC
simulation \cite{pelagos-qsw2026}. Estimating logical error rates requires
decoding many independent syndromes for the same circuit. P{\^{e}}lagos
processes these syndromes in batches so that GPU kernels perform the same
sparse graph operations across multiple decoding instances. The graph
structure is shared across the batch, while each syndrome has its own
messages, marginals, and convergence state. The implementation uses Triton
to compile the kernels for different GPU platforms
\cite{pelagos-qsw2026,tillet2019triton}.

Each BP iteration computes check-to-variable messages, updates
variable-to-check messages and marginals, tests syndrome agreement, and
records valid candidates. The check update finds the two smallest incoming
message magnitudes and their sign parity, allowing outgoing messages to
exclude their destination edge. The variable update fuses memory mixing,
message accumulation, marginal updates, and hard decisions. Candidate
selection and convergence flags remain on the GPU; batch lanes that have
found the requested number of solutions become inactive. Keeping this state
on the device reduces host synchronization and avoids transferring
intermediate decoding arrays after each iteration \cite{pelagos-qsw2026}.

The sparse message arrays dominate the decoder's data movement.
P{\^{e}}lagos supports both a syndrome-major layout $[B,E]$ and an
edge-major layout $[E,B]$, where $B$ is the batch size and $E$ is the
number of graph edges. The latter groups batch lanes for coalesced access,
while the former can benefit from cache reuse on smaller workloads.
Autotuning selects the layout and kernel configuration for the workload and
device. A persistent configuration cache records these choices together
with hardware and software fingerprints \cite{pelagos-qsw2026}.

For multiple GPUs, the existing design partitions the syndrome batch into
independent shards. Each GPU holds a graph replica and runs its own decode
loop, with no message exchange between GPUs. Corrections are gathered after
decoding; Ray distributes work across nodes for larger simulation studies
\cite{pelagos-qsw2026}. This decomposition provides the basis for our study.
We retain the single-GPU kernels and layouts and optimize the surrounding
node-local preprocessing, process ownership, and output collection.

\section{Multi-GPU Orchestration}
\label{sec:orchestration}
\subsection{Context Ownership}
\label{sec:context-ownership}
The original multi-GPU path constructs multiple GPU decoders in one Python
process and dispatches them through host threads. This organization keeps batch partitioning simple, but the parent process owns every HIP context. Process isolation therefore requires workers to establish device ownership before any process constructs a GPU decoder.

Our process-isolated design gives each physical GPU one persistent worker process. The GPU runtime fixes its visible devices at initialization, so each worker sets its assignment before that point. It then sees one physical GPU, addresses it as a single logical device, and runs the unmodified single-GPU decoder. Workers start as fresh Python interpreters and initialize the GPU runtime after setting their device assignment. The parent process holds no GPU context. Each worker also keeps a private compiler cache and, during tuning, a private measurement log.

\begin{figure*}[t]
  \centering
  \includegraphics[width=\textwidth]{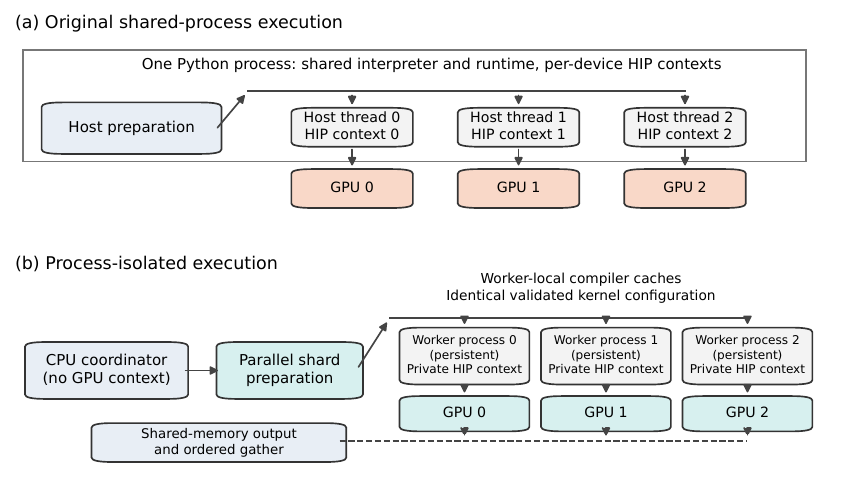}
  \caption{Orchestration before and after process isolation. The diagram shows
  three workers for clarity; the implementation supports up to eight.
  The baseline manages all device contexts within one host process;
  the optimized design isolates each device context in a separate worker process.}
  \label{fig:orchestration}
\end{figure*}

Figure~\ref{fig:orchestration} contrasts the original and process-isolated
execution paths. Solid arrows show control and input flow, and dashed arrows show the shared-memory output path.

\subsection{Parallel Preprocessing and Shared Output}
\label{sec:host-path}
Before decoding, the CPU computes the syndrome and observable values associated with each sampled error. The original path preprocesses the total batch serially. Our implementation splits the batch into one contiguous range per GPU, computes the independent sparse operations for those ranges in a host thread pool, and concatenates the resulting arrays once before dispatch. Each worker then decodes the range prepared for it.

Large correction arrays after decoding create a second host-side cost. Returning those arrays through process pipes requires serialization and an additional copy for every worker. We allocate one shared-memory array with the shape of the global correction output and assign each worker its destination slice. A worker copies its device result directly into that slice and returns smaller metadata through its pipe. The coordinator materializes the final CPU tensor from the shared array after all workers finish. This design preserves the decoder interface while removing the correction-array payload from the process pipes.

\subsection{Configuration Control}
\label{sec:configuration-control}
Our scaling experiments preserve the 1,024-syndrome per-GPU problem shape at every GPU count. We therefore pin one validated, system- and circuit-specific configuration cache for each shape rather than retuning after each change in GPU count, $G$. Every worker for a given system and circuit receives the same cache and executes the same kernel mapping.
Section~\ref{sec:accuracy} reports the logical error rate separately as a non-comparative sanity check.

\section{Results}
\label{sec:results}
\subsection{Evaluation Methodology}
\label{sec:methodology}
\subsubsection{Workload and Systems}
\label{sec:workload-platforms}
We evaluate decoding of the one-gross $[[144,12,12]]$ and two-gross $[[288,12,18]]$ bivariate bicycle code~\cite{bravyi2024bb} memory experiments using correlated XYZ decoding. We use a linear circuit-level noise model with physical error rate $p=0.003$: each nonidentity single-qubit Pauli error occurs with probability $p/3$, each nonidentity two-qubit Pauli error with probability $p/15$, and state-preparation and measurement errors with probability $p$~\cite{muller2025relay}.
Each circuit uses $d$ noisy syndrome-extraction rounds followed by one noiseless round, where $d$ is the code distance, as in~\cite{muller2025relay}.
The one-gross detector error model (DEM) contains 1,872 checks, 71,280 error mechanisms, and 419,904 nonzero check-matrix entries across 13 detector layers (12 noisy rounds plus one noiseless round). The two-gross DEM contains 5,472 checks, 213,408 error mechanisms, and 1,257,984 nonzero entries across 19 detector layers (18 noisy rounds plus one noiseless round).

Table~\ref{tab:decoder-params} lists the Relay-BP parameters and the resulting per-circuit decoding work. We take every algorithmic parameter from the P{\^{e}}lagos paper without modification, so the values match the published configuration \cite{pelagos-qsw2026,muller2025relay}, and they remain fixed across systems and GPU counts.
The implementation is available in the public P{\^{e}}lagos
repository at commit
\href{https://github.com/quantum-supercomputing/pelagos/commit/0127224e02ee4f7e4a43e7f038269509d780990e}{\texttt{0127224}}.\footnote{\url{https://github.com/quantum-supercomputing/pelagos}}
Our harness overrides only the execution settings that scaling requires: it selects throughput mode, sets the logical GPU list, and scales the batch with GPU count. We also evaluate the single relay depth $R=50$ rather than the published depth ladder.

\begin{table}[t]
  \centering
  \caption{Relay-BP parameters and decoding work. Iteration counts and logical
  error rates are single-GPU medians at $R=50$; ranges span both platforms.}
  \label{tab:decoder-params}
  \resizebox{\columnwidth}{!}{\begin{tabular}{@{}lll@{}}
    \toprule
    Parameter & $[[144,12,12]]$ & $[[288,12,18]]$ \\
    \midrule
    \multicolumn{3}{@{}l}{\emph{Algorithmic parameters}} \\
    Pre-relay BP iterations & \multicolumn{2}{c}{80} \\
    Iterations per relay set & \multicolumn{2}{c}{60} \\
    Relay sets $R$ & \multicolumn{2}{c}{50} \\
    Convergence count $S$ & \multicolumn{2}{c}{1} \\
    Normalized min-sum scaling $\alpha$ & \multicolumn{2}{c}{1} \\
    Initial memory strength $\gamma_0$ & \multicolumn{2}{c}{0.125} \\
    Memory-strength interval & $[-0.24,\,0.66]$ & $[-0.161,\,0.815]$ \\
    Message precision & \multicolumn{2}{c}{\texttt{float32}} \\
    Message schedule & \multicolumn{2}{c}{flood-fill} \\
    Physical error rate $p$ & \multicolumn{2}{c}{0.003} \\
    Relay seed & \multicolumn{2}{c}{0} \\
    Syndromes per GPU & \multicolumn{2}{c}{1,024} \\
    \midrule
    \multicolumn{3}{@{}l}{\emph{Per-circuit model size}} \\
    Noisy syndrome-extraction rounds & 12 & 18 \\
    Checks & 1,872 & 5,472 \\
    Error mechanisms & 71,280 & 213,408 \\
    Check-matrix nonzeros & 419,904 & 1,257,984 \\
    \midrule
    \multicolumn{3}{@{}l}{\emph{Measured decoding work}} \\
    BP iterations per syndrome & 77.8--78.9 & 106.2--107.9 \\
    Average relay legs & 1.4 & 1.6 \\
    Relative edge updates per syndrome & $1.0\times$ & $4.1\times$ \\
    Logical error rate & $2.93\times10^{-4}$ & 0 observed \\
    \bottomrule
  \end{tabular}
  }
\end{table}

Table~\ref{tab:platforms} summarizes the evaluated node configurations. The
MI355X configuration uses the Lux AI + HPC  supercomputer that is deployed at the Oak Ridge National Laboratory
~\cite{olcf-lux}. Both configurations run PyTorch atop
\mbox{AMD ROCm\texttrademark{} software}. Both systems use the same P{\^{e}}lagos source revision and matching Python package versions. The capacity and
bandwidth values come from the product specifications
\cite{amd-mi300x-spec,amd-mi355x-spec}, while each job records its host
processor, core count, and memory capacity.
\begin{table}[t]
  \centering
  \caption{Hardware and software setup. Memory capacity and peak bandwidth
  are specified per GPU.}
  \label{tab:platforms}
  \resizebox{\columnwidth}{!}{\begin{tabular}{@{}lll@{}}
    \toprule
    & MI300X & MI355X \\
    \midrule
    GPU & AMD Instinct MI300X GPU & AMD Instinct MI355X GPU \\
    Architecture & \mbox{AMD CDNA\texttrademark{} 3 architecture}
                 & AMD CDNA 4 architecture \\
    Compute units$^{\mathrm{a}}$ & 304 & 256 \\
    GPU memory & 192\,GB HBM3 & 288\,GB HBM3E \\
    Peak bandwidth & 5.3\,TB/s & 8.0\,TB/s \\
    GPUs per node & 8 & 8 \\
    Host processor & \mbox{AMD EPYC\texttrademark{} 9684X processor}
                   & AMD EPYC 9575F processor \\
    Host cores & $2\times96$ & $2\times64$ \\
    Host memory & 2.1\,TiB & 2.9\,TiB \\
    \midrule
    PyTorch & \multicolumn{2}{c}{2.13.0+rocm7.2} \\
    Triton & \multicolumn{2}{c}{3.7.1} \\
    \bottomrule
  \end{tabular}
  }
  \par\smallskip
  {\footnotesize\raggedright $^{\mathrm{a}}$MI355X has 256 compute units
  (CDNA 4), compared with 304 on MI300X (CDNA 3), but provides higher peak
  memory bandwidth (8.0 versus 5.3\,TB/s). Compute-unit counts across these
  architectures alone do not determine application throughput
  \cite{amd-mi300x-spec,amd-mi355x-spec}.\par}
\end{table}

\subsubsection{Scaling Metrics}
\label{sec:scaling-definitions}
We report three throughput series:
\begin{itemize}
  \item \textbf{P{\^{e}}lagos-Baseline} is the original decoder, which drives all GPUs
  from one interpreter using host threads and reports end-to-end throughput.
  \item \textbf{P{\^{e}}lagos-E2E} enables the full process-isolated design
  and also reports throughput calculated using the end-to-end time.
  \item \textbf{P{\^{e}}lagos-Parallel} uses the same P{\^{e}}lagos-E2E runs but calculates throughput using GPU decoding time, defined as the maximum elapsed worker decode time. This timing
  excludes coordinator preprocessing, dispatch, and subsequent result handling.
\end{itemize}

Our scaling experiments fix the per-GPU batch size at 1,024 syndromes while
increasing the total batch from 1,024 to 8,192 for eight GPUs within one node.
This fixed-work-per-GPU setup measures whether aggregate throughput grows
with GPU count after including orchestration and result handling, consistent
with scaled-work analysis \cite{gustafson1988reevaluating}.

Specifically, for GPU count $G$ and decoder variant $v$, the scaling speedup is
\begin{equation}
S_v(G)=\frac{T_v(G)}{T_v(1)},
\label{eq:speedup}
\end{equation}
where $T_v(G)$ is the median decoding iteration throughput of
P{\^{e}}lagos-Baseline, P{\^{e}}lagos-E2E, or P{\^{e}}lagos-Parallel on $G$
GPUs. Each variant uses its own one-GPU throughput for the same circuit and
system.
To compare throughput across platforms without mixing normalization baselines, we use the ratio
\begin{equation}
\rho(G)=\frac{T_{\mathrm{E2E,MI355X}}(G)}
{T_{\mathrm{E2E,MI300X}}(G)}.
\label{eq:platform-ratio}
\end{equation}
The ratio $\rho(G)$ uses belief-propagation iterations per second to compare
P{\^{e}}lagos-E2E between systems at the same GPU count.
We report throughput in both belief-propagation iterations per second and
decodes per second (syndromes per second).

\subsubsection{Experimental Setup}
\label{sec:experimental-setup}
We allocate complete nodes exclusively and test both circuits at $G\in\{1,2,4,8\}$. Each worker decodes a contiguous index range, and the coordinator reassembles the results in the original input order. For each point, we run one warm-up (excluded from the results) followed by three measured warm-cache repetitions. The workers persist across all four runs, so kernel loading and allocator warm-up occur only in the initial warm-up run rather than inside a measured decode. We report the median throughput across the three measured repetitions. The per-GPU batch remains 1,024, so the total batch ranges from 1,024 to 8,192 syndromes. Our throughput metric includes host preprocessing, dispatch, GPU execution, and output collection, so it measures the complete decoding workflow rather than GPU kernels alone.

\subsection{End-to-End Scaling}
\label{sec:final-scaling}
\begin{figure}[tb]
  \centering
  \includegraphics[width=\columnwidth]{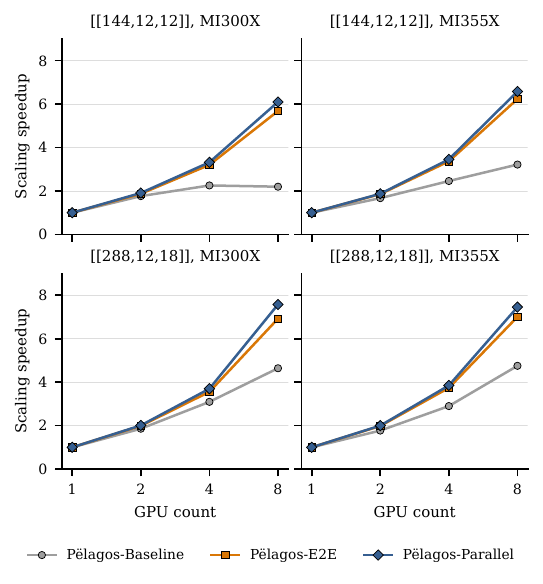}
  \caption{Scaling speedup at $R=50$ with 1,024 syndromes per GPU. Each series
  is normalized to its own one-GPU throughput for the same circuit and system.}
  \label{fig:scaling-attribution}
\end{figure}

Figure~\ref{fig:scaling-attribution} plots the three series for both circuits
on both systems. Table~\ref{tab:orchestration-gain} reports absolute throughput,
matched-GPU gain, and the MI355X/MI300X ratio. The original P{\^{e}}lagos
study reports that P{\^{e}}lagos-Baseline has higher throughput than CUDA-Q QEC
on MI300X \cite{pelagos-qsw2026}. In this paper, we compare our optimized
orchestration against P{\^{e}}lagos-Baseline under the same experimental
settings. From the figure, one can make the following observations.
\begin{table*}[t]
  \centering
  \caption{Throughput at each GPU count. Parentheses show E2E/Baseline gains;
  $\rho(G)$ is the MI355X/MI300X E2E throughput ratio.}
  \label{tab:platform-ratio}
  \label{tab:orchestration-gain}
  \resizebox{\textwidth}{!}{\begin{tabular}{llrrrrrrrrr}
    \toprule
    & & \multicolumn{4}{c}{MI300X} & \multicolumn{4}{c}{MI355X} & \\
    \cmidrule(lr){3-6}\cmidrule(lr){7-10}
    & & \multicolumn{2}{c}{P{\^{e}}lagos-Baseline}
        & \multicolumn{2}{c}{P{\^{e}}lagos-E2E}
        & \multicolumn{2}{c}{P{\^{e}}lagos-Baseline}
        & \multicolumn{2}{c}{P{\^{e}}lagos-E2E} & \\
    \cmidrule(lr){3-4}\cmidrule(lr){5-6}\cmidrule(lr){7-8}\cmidrule(lr){9-10}
    Code & GPUs & Iterations/s & Decodes/s & Iterations/s & Decodes/s
         & Iterations/s & Decodes/s & Iterations/s & Decodes/s & $\rho(G)$ \\
    \midrule
    $[[144,12,12]]$ & 1 & 36,206 & 465 & 36,938 ($1.02\times$) & 475 & 45,602 & 578 & 48,648 ($1.07\times$) & 617 & $1.32\times$ \\
                     & 2 & 63,836 & 819 & 69,086 ($1.08\times$) & 886 & 76,141 & 983 & 90,205 ($1.18\times$) & 1,166 & $1.31\times$ \\
                     & 4 & 81,598 & 1,044 & 118,265 ($1.45\times$) & 1,513 & 112,065 & 1,441 & 163,740 ($1.46\times$) & 2,109 & $1.38\times$ \\
                     & 8 & 79,520 & 1,019 & 209,826 ($2.64\times$) & 2,686 & 146,639 & 1,864 & 303,185 ($2.07\times$) & 3,845 & $1.44\times$ \\
    \midrule
    $[[288,12,18]]$ & 1 & 14,976 & 141 & 14,810 ($0.99\times$) & 139 & 20,706 & 192 & 19,165 ($0.93\times$) & 178 & $1.29\times$ \\
                     & 2 & 27,804 & 261 & 29,434 ($1.06\times$) & 276 & 36,676 & 342 & 37,981 ($1.04\times$) & 354 & $1.29\times$ \\
                     & 4 & 46,326 & 435 & 52,643 ($1.14\times$) & 494 & 60,019 & 561 & 71,623 ($1.19\times$) & 670 & $1.36\times$ \\
                     & 8 & 69,485 & 658 & 102,521 ($1.48\times$) & 970 & 98,389 & 928 & 134,282 ($1.36\times$) & 1,265 & $1.31\times$ \\
    \bottomrule
  \end{tabular}
  }
\end{table*}

\begin{figure}[tb]
  \centering
  \includegraphics[width=\columnwidth]{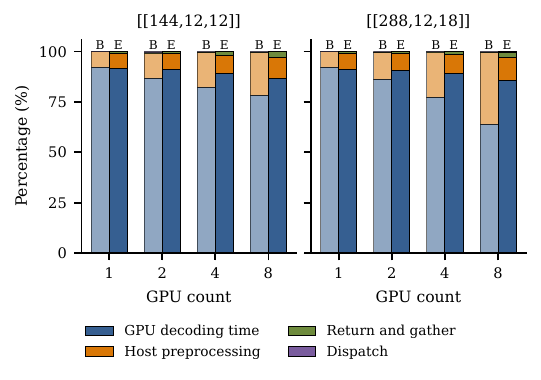}
  \caption{Matched decode-time composition on MI355X at $R=50$. Within each
  GPU-count group, B denotes P{\^{e}}lagos-Baseline and E denotes
  P{\^{e}}lagos-E2E.}
  \label{fig:time-decomposition}
\end{figure}

First, P{\^{e}}lagos-E2E scales better than P{\^{e}}lagos-Baseline on both
circuits and systems. For one-gross, the eight-GPU scaling speedup increases
from $2.20\times$ to $5.68\times$ on MI300X and from $3.22\times$ to
$6.23\times$ on MI355X. For two-gross, it increases from $4.64\times$ to
$6.92\times$ and from $4.75\times$ to $7.01\times$, respectively.
Figure~\ref{fig:time-decomposition} helps explain this improvement by
comparing the execution-time composition before and after optimization on
MI355X. As GPU count grows, the total batch increases. Because the baseline
performs preprocessing  serially, CPU preprocessing time grows roughly
linearly with GPU count. Our design instead prepares the fixed-size shards
concurrently, keeping preprocessing time relatively stable as more GPUs
are added (Section~\ref{sec:host-path}). On MI355X, CPU
preprocessing time per batch increases from 0.129 to
0.974\,s for one-gross and from 0.437 to 3.213\,s for two-gross when scaling
from one to eight GPUs. After optimization, the corresponding increases
are only 0.132 to 0.222\,s and 0.446 to 0.733\,s. These absolute times
complement the fractions shown in Figure~\ref{fig:time-decomposition}:
parallel preprocessing limits the growth of CPU work on the critical path,
helping P{\^{e}}lagos-E2E sustain better scaling than the baseline.

Second, P{\^{e}}lagos-Parallel scales effectively on both systems, with near-linear scaling for the two-gross circuit. At eight GPUs, it reaches
$6.10\times$ on MI300X and $6.58\times$ on MI355X for the one-gross circuit,
and $7.58\times$ and $7.45\times$ for the two-gross circuit.
This scaling benefits from the independence of syndrome shards: each GPU
executes the decoder on its own shard without exchanging intermediate messages
with other GPUs. Since the per-GPU batch remains fixed, adding GPUs increases
the number of syndromes processed concurrently without enlarging each worker's
batch. However, equal batch sizes do not guarantee equal decoding times because
syndromes can require different numbers of BP iterations and relay legs. The
GPU decoding time is determined by the slowest worker. The measured
results show that this independent-shard design scales effectively on both
systems, with the two-gross circuit approaching the ideal eightfold throughput
increase.

Third, MI355X sustains higher P{\^{e}}lagos-E2E throughput than MI300X for both
circuits at every tested GPU count. For one-gross, MI355X delivers $1.32\times$
the throughput of MI300X at one GPU and $1.44\times$ at eight GPUs. For
two-gross, the corresponding improvements are $1.29\times$ and $1.31\times$
(Table~\ref{tab:orchestration-gain}). One reason for this advantage is MI355X's
higher memory bandwidth: it provides $8.0$\,TB/s per GPU, compared with
$5.3$\,TB/s on MI300X (Table~\ref{tab:platforms}). Each BP iteration reads
incoming messages along the decoding graph's edges and writes updated messages
and marginals. These operations repeatedly move large sparse arrays through GPU
memory. Higher memory bandwidth allows this data movement to complete faster,
improving decoding throughput.

\begin{figure*}[!t]
  \centering
  \includegraphics[width=\textwidth]{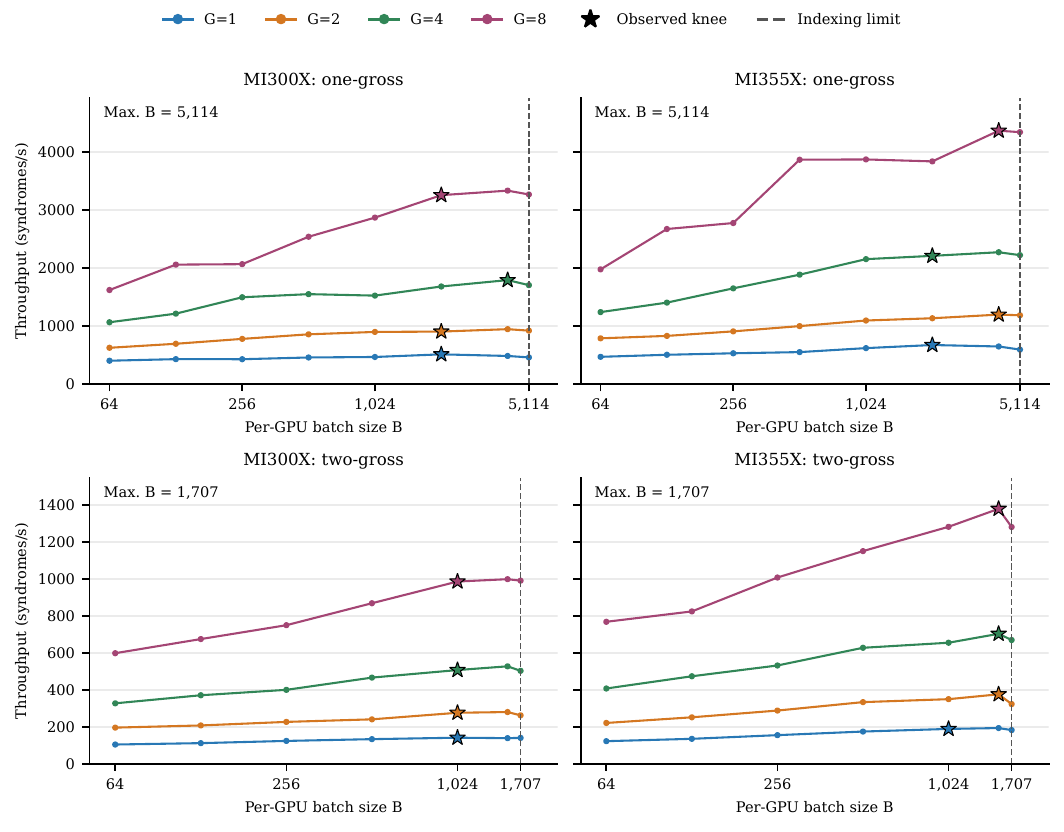}
  \caption{P{\^{e}}lagos-E2E throughput versus per-GPU batch size at
  $R=50$. Points show medians of three measured repetitions.
  Stars mark $B^*(G)$, the smallest tested batch
  whose median throughput is within 5\% of the best measured median for
  that GPU count, circuit, and platform. Dashed lines mark the
  largest supported batches imposed by indexing.}
  \label{fig:batch-sweep}
\end{figure*}

\subsection{Ablation and Microarchitecture Analysis}
\label{sec:microarchitecture}
Increasing the per-GPU batch size improves throughput over small batches. We sweep batch size $B$ at GPU count $G\in\{1,2,4,8\}$ on both platforms at $R=50$, using P{\^{e}}lagos-E2E (Figure~\ref{fig:batch-sweep}). 
The tested batch sizes are 64, 128, 256, 512, and 1,024 for both circuits, followed by 2,048, 4,096, and 5,114 for one-gross, and 1,536 and 1,707 for two-gross. 
Note that the maximum batch size is bounded by indexing rather than HBM capacity. Specifically, signed 32-bit indexing requires $E\times B\leq2^{31}-1$, where $E$ is the number of edges in each circuit's Tanner graph. This limits $B$ to 5,114 for one-gross and 1,707 for two-gross on both platforms.

In Figure~\ref{fig:batch-sweep}, we report the median throughput across three measured repetitions after warm-up. For a fair comparison across different batch sizes, each circuit uses the configurations autotuned at $B=1{,}024$ for all five decoding kernels across all batch sizes and GPU counts.
It is worth noting that autotuning the kernel configurations at each batch size independently may yield different throughput.
This sweep evaluates how batch size affects P{\^{e}}lagos-E2E throughput at each GPU count.

We define the observed knee $B^*(G)$ as the smallest tested batch size that attains at least 95\% of the maximum median throughput under the same circuit, platform, and GPU count.
The knee points are marked with stars in Figure~\ref{fig:batch-sweep}. On MI300X, the one-gross knees are $2{,}048$, $2{,}048$, $4{,}096$, and $2{,}048$ for one, two, four, and eight GPUs, respectively. The two-gross knee is $1{,}024$ at all four GPU counts.
On MI355X, the one-gross knees are $2{,}048$, $4{,}096$, $2{,}048$, and $4{,}096$ for one, two, four, and eight GPUs, respectively. For two-gross, the knee is $1{,}024$ on one GPU and $1{,}536$ on two, four, and eight GPUs.
On MI355X at $B=1{,}024$, increasing from four to eight GPUs raises throughput from 2,153 to 3,871 syndromes/s for one-gross ($1.80\times$), and from 656 to 1,283 syndromes/s for two-gross ($1.96\times$).
Overall, increasing batch size improves throughput over small batches, with diminishing gains on several curves. The potential reasons are i) increasing CPU processing time for larger batches and ii) GPU memory-bandwidth saturation, which would limit further throughput gains.

We also conduct a microarchitecture analysis by profiling both circuits on a single MI355X GPU. Larger batches increase measured GPU occupancy while reducing cache reuse (Table~\ref{tab:memory-profile}).
Specifically, increasing $B$ provides more independent syndrome work, but also enlarges the edge-message arrays of shape $[E,B]$. For one-gross, occupancy rises from 30.9\% at $B=64$ to 66.3\% at $B=1{,}024$ and 80.8\% at $B=5{,}114$. For two-gross, occupancy increases from 35.8\% to 70.9\% across the tested sizes. Meanwhile, from the smallest to the largest tested batch, L2 hit rates decrease from 57.5\% to 42.9\% for one-gross and from 65.2\% to 52.5\% for two-gross. Vector-L1 hit rates also decrease. Thus, larger batches expose more parallel work with lower cache-hit rates.

\begin{table}[tb]
  \centering
  \caption{Single-GPU MI355X memory traffic, cache hit rates, and occupancy.}
  \label{tab:memory-profile}
  \small
  \setlength{\tabcolsep}{3pt}
  \begin{tabular}{lrrrrr}
    \toprule
    Circuit & Batch & Traffic$^{\mathrm{a}}$ & L2 hit$^{\mathrm{b}}$ & vL1 hit$^{\mathrm{c}}$ & Occupancy$^{\mathrm{d}}$ \\
            & size & (TB/s) & (\%) & (\%) & (\%) \\
    \midrule
    One-gross & 64 & 2.30 & 57.5 & 61.1 & 30.9 \\
              & 1,024 & 3.45 & 44.9 & 57.1 & 66.3 \\
              & 5,114 & 3.84 & 42.9 & 49.5 & 80.8 \\
    Two-gross & 64 & 1.57 & 65.2 & 61.5 & 35.8 \\
              & 1,024 & 3.03 & 54.1 & 55.1 & 66.9 \\
              & 1,707 & 3.26 & 52.5 & 53.0 & 70.9 \\
    \bottomrule
  \end{tabular}
  \par\smallskip
  \begin{minipage}{\columnwidth}
    \footnotesize\raggedright
    $^{\mathrm{a}}$Traffic: average requested HBM bandwidth, calculated as
    HBM-directed bytes at the L2 interface divided by total selected-kernel
    execution time; measured on Lux with ROCm SDK.\par
    $^{\mathrm{b}}$L2 hit: request-weighted L2 cache hit rate, measured with
    the same profiler on Lux.\par
    $^{\mathrm{c}}$vL1 hit: request-weighted vector-L1 cache hit rate,
    measured with the same profiler on Lux.\par
    $^{\mathrm{d}}$Occupancy: ROCm SDK active-cycle-normalized metric measured
    on Lux, duration-weighted over measured decoder kernels after warm-up
    and calibration, and averaged across two runs. Short-kernel measurements
    are subject to profiler accuracy limitations.
  \end{minipage}
\end{table}

\subsection{Accuracy Checks}
\label{sec:accuracy}
We use logical error rate (LER) as a correctness check rather than a scaling
metric. Each worker decodes an independent shard, so the multi-GPU path performs
no cross-GPU floating-point reduction. However, worker-local relay-memory
sampling and the growing syndrome population mean that different GPU counts
need not process identical decoder trajectories. For $[[144,12,12]]$, the
observed LER ranges from $8.14\times10^{-5}$ to $3.91\times10^{-4}$; the small
failure counts do not support an accuracy comparison across GPU counts. After
correcting the two-gross memory-strength interval to $[-0.161,0.815]$, we
observe no logical failures in any measured run at $R=50$. The two-gross
circuit requires an average of $1.6$ relay legs per syndrome. The published
P{\^{e}}lagos study provides the broader decoder-quality evaluation
\cite{pelagos-qsw2026}.

\section{Related Work}
\label{sec:related-work}
M{\"u}ller et al.\ introduced Relay-BP as a memory-assisted decoder for quantum
codes \cite{muller2025relay}. Diamantopoulos et al.\ mapped Relay-BP to GPU kernels
in P{\^{e}}lagos and evaluated its layouts, portability, decoder quality, and
large-cluster deployment \cite{pelagos-qsw2026}. We build directly on that
implementation and focus on process-isolated host orchestration rather than
cluster-scale deployment. CUDA-Q QEC provides a GPU-accelerated quantum
low-density parity-check decoder that also implements belief propagation,
ordered-statistics post-processing, and Relay-BP \cite{cudaq-qec}.
The
published P{\^{e}}lagos study reports a comparison against CUDA-Q QEC on
MI300X \cite{pelagos-qsw2026}. We do
not repeat that comparison, and instead hold the decoder configuration fixed while varying
only the orchestration.

General autotuning systems, such as OpenTuner and the tensor virtual machine
(TVM) compiler, search implementation spaces using measured objectives
\cite{ansel2014opentuner,chen2018tvm}.
Triton exposes tile, warp, and pipeline parameters for generated GPU kernels
\cite{tillet2019triton}. P{\^{e}}lagos uses target-specific kernel
configurations, and our scaling design reuses one validated configuration per
system because the per-GPU problem shape remains fixed.

Scaled-work analysis evaluates throughput as the problem size grows with the
available processors \cite{gustafson1988reevaluating}. Relay-BP adds a practical
complication: data-dependent convergence changes active work across shards even
when every GPU receives the same number of syndromes. Our evaluation measures
this scaling behavior at $R=50$.

\section{Conclusion}
\label{sec:conclusion}
We present a process-isolated orchestration design for accelerating large-scale
QEC simulation with the P{\^{e}}lagos Relay-BP decoder within an eight-GPU node.
The design assigns one persistent process and one private HIP context to each
GPU, prepares shards concurrently, and returns large
outputs through shared memory. We evaluate the one-gross and two-gross circuits
on one, two, four, and eight GPUs while retaining a batch of 1,024 syndromes
per GPU.

At $R=50$, eight-GPU P{\^{e}}lagos-E2E speedup ranges from $5.68\times$ to
$6.92\times$ on MI300X and from $6.23\times$ to $7.01\times$ on MI355X across the two
circuits. Against the original decoder, the design more than doubles eight-GPU
throughput on the one-gross circuit and improves two-gross throughput by
$1.48\times$ on MI300X and $1.36\times$ on MI355X. These results demonstrate
that process-isolated orchestration increases the throughput available for
logical-error-rate estimation and decoder-parameter exploration, supporting
the large-scale evaluations needed to develop effective QEC decoders.

\section*{Acknowledgment}
IBM, the IBM logo, and ibm.com® are trademarks or registered trademarks of International Business Machines Corporation, in the United States and/or other countries. A current list of IBM trademarks is available on http://www.ibm.com/trademark.

AMD, the AMD Arrow logo, AMD CDNA, AMD EPYC, AMD Instinct, AMD ROCm, and
combinations thereof are trademarks of Advanced Micro Devices, Inc. Other
product names used in this publication are for identification purposes only
and may be trademarks of their respective companies.

\bibliographystyle{IEEEtran}
\bibliography{references}

\begin{thebibliography}{10}
\providecommand{\url}[1]{#1}
\csname url@samestyle\endcsname
\providecommand{\newblock}{\relax}
\providecommand{\bibinfo}[2]{#2}
\providecommand{\BIBentrySTDinterwordspacing}{\spaceskip=0pt\relax}
\providecommand{\BIBentryALTinterwordstretchfactor}{4}
\providecommand{\BIBentryALTinterwordspacing}{\spaceskip=\fontdimen2\font plus
\BIBentryALTinterwordstretchfactor\fontdimen3\font minus \fontdimen4\font\relax}
\providecommand{\BIBforeignlanguage}[2]{{%
\expandafter\ifx\csname l@#1\endcsname\relax
\typeout{** WARNING: IEEEtran.bst: No hyphenation pattern has been}%
\typeout{** loaded for the language `#1'. Using the pattern for}%
\typeout{** the default language instead.}%
\else
\language=\csname l@#1\endcsname
\fi
#2}}
\providecommand{\BIBdecl}{\relax}
\BIBdecl

\bibitem{beverland2025failfast}
\BIBentryALTinterwordspacing
M.~E. Beverland, M.~Carroll, A.~W. Cross, and T.~J. Yoder, ``Fail fast: Techniques to probe rare events in quantum error correction,'' arXiv:2511.15177, 2025. [Online]. Available: \url{https://arxiv.org/abs/2511.15177}
\BIBentrySTDinterwordspacing

\bibitem{bravyi2024bb}
S.~Bravyi, A.~W. Cross, J.~M. Gambetta, D.~Maslov, P.~Rall, and T.~J. Yoder, ``High-threshold and low-overhead fault-tolerant quantum memory,'' \emph{Nature}, vol. 627, no. 8005, pp. 778--782, Mar. 2024.

\bibitem{muller2025relay}
\BIBentryALTinterwordspacing
T.~M{\"u}ller, T.~Alexander, M.~E. Beverland, M.~B{\"u}hler, B.~R. Johnson, T.~Maurer, and D.~Vandeth, ``Improved belief propagation is sufficient for real-time decoding of quantum memory,'' arXiv:2506.01779, 2025. [Online]. Available: \url{https://arxiv.org/abs/2506.01779}
\BIBentrySTDinterwordspacing

\bibitem{mayer2025rare}
\BIBentryALTinterwordspacing
C.~Mayer, A.~Ganti, U.~Onunkwo, T.~Metodi, B.~Anker, and J.~Skryzalin, ``Rare event simulation of quantum error-correcting circuits,'' arXiv:2509.13678, 2025. [Online]. Available: \url{https://arxiv.org/abs/2509.13678}
\BIBentrySTDinterwordspacing

\bibitem{bravyi2013rare}
S.~Bravyi and A.~Vargo, ``Simulation of rare events in quantum error correction,'' \emph{Phys. Rev. A}, vol.~88, no.~6, p. 062308, 2013.

\bibitem{yoder2025tourdegross}
\BIBentryALTinterwordspacing
T.~J. Yoder, E.~Schoute, P.~Rall, E.~Pritchett, J.~M. Gambetta, A.~W. Cross, M.~Carroll, and M.~E. Beverland, ``Tour de gross: A modular quantum computer based on bivariate bicycle codes,'' arXiv:2506.03094, 2025. [Online]. Available: \url{https://arxiv.org/abs/2506.03094}
\BIBentrySTDinterwordspacing

\bibitem{pelagos-qsw2026}
D.~Diamantopoulos, L.~von Niederh{\"a}usern, A.~S. Anghel, R.~Pletka, A.~Kayi, T.~M{\"u}ller, N.~Papandreou, J.~Blanusa, S.~Sarafijanovic, N.~M. Gonzalez, F.~Abel, C.~Hagleitner, S.~Seelam, B.~Johnson, R.~Haas, and H.~Pozidis, ``P{\^{e}}lagos: An adaptive {GPU} decoder for error correction simulations of quantum {LDPC} codes,'' in \emph{2026 IEEE International Conference on Quantum Software (QSW)}, 2026, pp. 152--165.

\bibitem{olcf-lux}
\BIBentryALTinterwordspacing
{Oak Ridge Leadership Computing Facility}, ``Lux: {ORNL}'s {AI} $+$ {HPC} supercomputer,'' accessed: Sep. 29, 2026. [Online]. Available: \url{https://www.olcf.ornl.gov/olcf-resources/compute-systems/lux/}
\BIBentrySTDinterwordspacing

\bibitem{amd-lux-science}
\BIBentryALTinterwordspacing
{AMD}, ``{AMD} accelerates scientific discovery: Powering next-gen supercomputers and leading {HPC} and {AI} innovation,'' 2025, accessed: Oct. 16, 2026. [Online]. Available: \url{https://www.amd.com/en/blogs/2025/amd-accelerates-scientific-discovery-powering-next-gen.html}
\BIBentrySTDinterwordspacing

\bibitem{Panteleev_2021}
\BIBentryALTinterwordspacing
P.~Panteleev and G.~Kalachev, ``Degenerate quantum {LDPC} codes with good finite length performance,'' \emph{Quantum}, vol.~5, p. 585, Nov. 2021. [Online]. Available: \url{http://dx.doi.org/10.22331/q-2021-11-22-585}
\BIBentrySTDinterwordspacing

\bibitem{Roffe2020}
\BIBentryALTinterwordspacing
J.~Roffe, D.~R. White, S.~Burton, and E.~Campbell, ``Decoding across the quantum low-density parity-check code landscape,'' \emph{Phys. Rev. Res.}, vol.~2, p. 043423, Dec 2020. [Online]. Available: \url{https://link.aps.org/doi/10.1103/PhysRevResearch.2.043423}
\BIBentrySTDinterwordspacing

\bibitem{tillet2019triton}
P.~Tillet, H.~T. Kung, and D.~Cox, ``Triton: An intermediate language and compiler for tiled neural network computations,'' in \emph{Proc. 3rd ACM SIGPLAN Int. Workshop Mach. Learn. Program. Lang.}, Jun. 2019, pp. 10--19.

\bibitem{amd-mi300x-spec}
\BIBentryALTinterwordspacing
{AMD}, ``{AMD Instinct MI300 Series Accelerators},'' accessed: Aug. 30, 2026. [Online]. Available: \url{https://www.amd.com/en/products/accelerators/instinct/mi300.html}
\BIBentrySTDinterwordspacing

\bibitem{amd-mi355x-spec}
\BIBentryALTinterwordspacing
{AMD}, ``{AMD Instinct MI355X GPUs},'' accessed: Aug. 30, 2026. [Online]. Available: \url{https://www.amd.com/en/products/accelerators/instinct/mi350/mi355x.html}
\BIBentrySTDinterwordspacing

\bibitem{gustafson1988reevaluating}
J.~L. Gustafson, ``Reevaluating {Amdahl}'s law,'' \emph{Commun. ACM}, vol.~31, no.~5, pp. 532--533, May 1988.

\bibitem{cudaq-qec}
\BIBentryALTinterwordspacing
{NVIDIA}, ``{QEC} decoders,'' {NVIDIA CUDA-Q QEC} documentation, accessed: Sep. 21, 2026. [Online]. Available: \url{https://nvidia.github.io/cudaq-qec/components/qec/decoders.html}
\BIBentrySTDinterwordspacing

\bibitem{ansel2014opentuner}
J.~Ansel, S.~Kamil, K.~Veeramachaneni, J.~Ragan-Kelley, J.~Bosboom, U.-M. O'Reilly, and S.~Amarasinghe, ``{OpenTuner}: An extensible framework for program autotuning,'' in \emph{Proc. 23rd Int. Conf. Parallel Archit. Compilation Techn.}, Aug. 2014, pp. 303--316.

\bibitem{chen2018tvm}
\BIBentryALTinterwordspacing
T.~Chen, T.~Moreau, Z.~Jiang, L.~Zheng, E.~Yan, H.~Shen, M.~Cowan, L.~Wang, Y.~Hu, L.~Ceze, C.~Guestrin, and A.~Krishnamurthy, ``{TVM}: An automated end-to-end optimizing compiler for deep learning,'' in \emph{Proc. 13th USENIX Symp. Oper. Syst. Design Implement.}\hskip 1em plus 0.5em minus 0.4em\relax Carlsbad, CA, USA: USENIX Association, Oct. 2018, pp. 578--594. [Online]. Available: \url{https://www.usenix.org/conference/osdi18/presentation/chen}
\BIBentrySTDinterwordspacing

\end{thebibliography}

\end{document}